\documentclass[12pt]{article}
\usepackage{graphics}
\usepackage{psfig}
\usepackage{amssymb}
\usepackage{amsmath}

\newcommand{\rf}[1]{(\ref{#1})}
\newcommand{\beq}{\begin{equation}}
\newcommand{\eeq}{\end{equation}}
\newcommand{\bea}{\begin{eqnarray}}
\newcommand{\eea}{\end{eqnarray}}

\newcommand{\e}{\mbox{e}}
\renewcommand{\d}{\mbox{d}}
\newcommand{\g}{\gamma}

\newcommand{\La}{\Lambda}
\renewcommand{\b}{\beta}

\renewcommand{\th}{\theta}
\newcommand{\Del}{\Delta}

\renewcommand{\k}{\kappa}

\newcommand{\oh}{\frac{1}{2}}

\newcommand{\ra}{\rangle}
\newcommand{\la}{\langle}
\newcommand{\prt}{\partial}
\newcommand{\mi}{\!-\!}
\newcommand{\equ}{\!=\!}
\newcommand{\pl}{\!+\!}

\newcommand{\cD}{{\cal D}}

\newcommand{\tL}{{\tilde{\La}}}
\newcommand{\tX}{{\tilde{X}}}

\newcommand{\bX}{{\bar{X}}}

\newcommand{\no}{\nonumber}

\newcommand{\SL}{\sqrt{\La}}

\begin{document}

\begin{center}
\vspace{24pt}
{ \large \bf The emergence of background geometry \\
from quantum fluctuations}

\vspace{30pt}

{\sl J. Ambj\o rn}$\,^{a,c}$
{\sl R. Janik}$\,^{b}$,
{\sl W. Westra}$\,^{c}$
and {\sl S. Zohren}$\,^{d}$

\vspace{48pt}
{\footnotesize

$^a$~The Niels Bohr Institute, Copenhagen University\\
Blegdamsvej 17, DK-2100 Copenhagen \O , Denmark.\\
{ email: ambjorn@nbi.dk}\\

\vspace{10pt}

$^b$~
Institute of Physics, Jagellonian University,\\
ul. Reymonta 4, 30-059 Krakow, Poland.\\
email: ufrjanik@if.uj.edu.pl\\

\vspace{10pt}

$^c$~Institute for Theoretical Physics, Utrecht University, \\
Leuvenlaan 4, NL-3584 CE Utrecht, The Netherlands.\\
email: w.westra@phys.uu.nl, ambjorn@phys.uu.nl\\

\vspace{10pt}

$^d$~Blackett Laboratory, Imperial College,\\
London SW7 2AZ, United Kingdom.\\
email: stefan.zohren@imperial.ac.uk\\
}
\vspace{96pt}

\end{center}


\begin{center}
{\bf Abstract}
\end{center}

We show how the quantization  of two-dimensional gravity
leads to an (Euclidean) quantum space-time where the average geometry
is that of constant negative curvature and where the Hartle-Hawking
boundary condition arises naturally.

\vspace{12pt}
\noindent


\newpage

\subsection*{Introduction}\label{intro}

Two-dimensional quantum gravity is not much of a gravity
theory in the sense that there are no propagating gravitons.
It has nevertheless been a fertile playground when it comes
to testing various aspects of diffeomorphism-invariant theories,
and it is potentially important for string theory which can be
viewed as two-dimensional quantum gravity coupled to specific,
conformal invariant matter fields. The 2d quantum gravity
aspect has been particularly important in the study of
non-critical string theories.

Most of the studies where the quantum gravity
aspect has been emphasized have considered
two-dimensional Euclidean quantum gravity with compact space-time.
The study of 2d Euclidean quantum gravity
with non-compact space-time
was initiated by the Zamolodchikovs (ZZ) \cite{zz} when they
showed how to use conformal bootstrap and the cluster-decomposition
properties to quantize Liouville theory on the pseudo-sphere
(the Poincare disk).

Martinec \cite{martinec} and Seiberg et al.\ \cite{ss}
showed how the work of ZZ fitted into framework of
non-critical string theory, where the ZZ-theory could
be reinterpreted as special branes, now called ZZ-branes.
Let $W_\tL(\tX)$ be the ordinary, so-called disk amplitude
for 2d Euclidean gravity on a compact space-time.
$\tX$ denotes the boundary cosmological constant of the disk and
$\tL$ the cosmological constant.
It was found that the ZZ-brane of 2d Euclidean gravity
was associated with the zero of
\beq\label{0.0}
W_\tL(\tX) = (\tX -\oh \sqrt{\tL})\sqrt{\tX + \sqrt{\tL}}.
\eeq
At first sight this is somewhat surprising since from a world-sheet
point of view the disk is compact while the Poincare disk
is non-compact. In \cite{aagk} it was shown how it could be understood
in terms of world sheet geometry, i.e.\ from a 2d quantum gravity
point of view: When the boundary cosmological constant $\tX$ reaches the value
$\tX \equ \sqrt{\tL}/2$ where the disk amplitude $W_\tL(\tX) \equ 0$,
the geodesic distance
from a generic point on the disk to the boundary diverges, in this
way effectively creating a non-compact space-time.

In this article we show that the same phenomenon occurs for
a different two-dimensional theory of quantum gravity called
quantum gravity from causal dynamical triangulations (short: CDT)
\cite{al}. This theory has been generalized to higher dimensions
where potentially interesting results have been obtained \cite{ajl5}
using computer simulations. However,
here we will concentrate on the two-dimensional theory which
can be solved analytically.

\section*{CDT}

The idea of CDT, i.e.\ quantum gravity defined via causal
dynamical triangulations, is two-fold: firstly, inspired by
Teitelboim \cite{teitelboim}, we insist, starting in a space-time with a
Lorentzian signature, that only causal histories contribute to
the quantum gravity path integral,
and secondly, we assume a global time-foliation.

``Dynamical triangulation'' (DT) provides a simple regularization
of the sum over {\it geometries} by providing a grid of
piecewise linear geometries constructed from building blocks
($d$-dimensional simplices if we want to construct a $d$-dimensional
geometry, see \cite{book,leshouches} for reviews).
The ultraviolet cut-off is the length of the side of the building blocks.
CDT uses DT as the regularization of the path integral
(see \cite{al,ajl5} for  detailed
descriptions of which causal geometries are included in the grid).

In two dimensions it is natural to study the  proper-time ``propagator'',
i.e.\ the amplitude for two space-like boundaries to be
separated a proper time (or geodesic distance) $T$. While this
is a somewhat special amplitude, it has the virtue that other
amplitudes, like the disk amplitude or the cylinder amplitude,
can be calculated if we know the proper-time propagator \cite{kawai0,kn,gk,al}.
When the path integral representation of this propagator
is defined using CDT we can further, for each causal piecewise
linear Lorentzian geometry, make an explicit rotation to a related
Euclidean geometry. After this rotation we perform 
the sum over geometries in the
this Euclidean regime.  This sum is now different from the full Euclidean
sum over geometries, leading to an alternative quantization
of 2d quantum gravity (CDT). Eventually we can perform a rotation back
from Euclidean proper time to Lorentzian proper time in the
propagator if needed.

In the following we will use continuum
notation. A derivation of the continuum expressions from the regularized
(lattice) expressions can be found in \cite{al}.
We assume space-time has the topology $S^1\times [0,1]$,
The action (rotated to Euclidean space-time) is:
\beq\label{2.a}
S[g] = \La \int \int \d x \d t \sqrt{g(x,t)} +
X \oint \d l_1 + Y \oint \d l_2,
\eeq
where $\La$ is  the cosmological constant, $X,Y$ are two boundary
cosmological constants, $g$ is a metric describing a geometry
of the kind mentioned above, and the line integrals refer to
the length of the boundaries, induced by $g$.
The propagator $G_\La(X,Y;T)$ is defined by
\beq\label{2.a0}
G_\La (X,Y;T) =
\int \cD [g] \; e^{-S[g]},
\eeq
where the functional integration is over all ``causal'' geometries $[g]$
such that the ``exit'' boundary with boundary cosmological constant
$Y$ is separated a geodesic distance $T$ from the ``entry'' boundary
with boundary cosmological constant $X$. As shown in \cite{al}, calculating
the path integral \rf{2.a0} using the CDT regularization and taking
the continuum limit where the side-length $a$ of the simplices goes
to zero leads to the following expression\footnote{The
asymmetry between $X$ and $Y$ is just due to the convention that
the entrance boundary contains a marked point. Symmetric expressions
where the the boundaries have no marked points or both have
marked points can be found in \cite{alnr}}:
\beq\label{2.a3}
G_\La (X,Y;T) = \frac{\bX^2(T,X)-\La}{X^2-\La} \; \frac{1}{\bX(T,X)+Y},
\eeq
where $\bX(T,X)$ is the solution of
\beq\label{2.3}
\frac{\d \bX}{\d T} = -(\bX^2-\La),~~~\bX(0,X)=X,
\eeq
or
\beq\label{2.4}
\bX(t,X)= \SL \coth \SL(t+t_0),~~~~X=\SL \coth \SL \,t_0.
\eeq
Viewing $G_\La(X,Y;T)$ as a propagator, 
$\bX(T)$ can be viewed as a ``runing'' boundary cosmological
constant, $T$ being the scale. If $X > - \SL$ then
$\bX(T) \to \SL$ for $T \to \infty$, $\SL$ being a ``fixed point''
(a zero of the ``$\b$-function'' $-(\bX^2-\La)$ in eq.\ \rf{2.3}).

Let $L_1$ denote the length of the entry boundary and $L_2$ the
length of the exit boundary. Rather than consider a situation
where the boundary cosmological constant $X$ is fixed we can
consider $L_1$ as fixed. We denote the corresponding propagator
$G_\La (L_1,Y;T)$. Similarly we can define $G_\La(X,L_2;T)$
and $G_\La(L_1,L_2;T)$. They are related by Laplace transformations.
For instance:
\beq\label{2.a5}
G_\La(X,Y;T)= \int_0^\infty \d L_2 \int_0^\infty \d L_1\;
G(L_1,L_2;T) \;\e^{-XL_1-YL_2}.
\eeq
and one has the following composition rule for the propagator:
\beq\label{2.a6}
G_\La (X,Y;T_1+T_2) = \int_0^\infty \d L \;
G_\La (X,L;T_1)\,G(L,Y,T_2).
\eeq

We can now calculate the expectation value of the length
of the spatial slice at  proper time $t \in [0,T]$:
\beq\label{2.a7}
\la L(t)\ra_{X,Y,T} =
\frac{1}{G_\La (X,Y;T)} \int_0^\infty \d L\;
G_\La (X,L;t) \;L\; G_\La (L,Y;T-t).
\eeq
In general there is no reason to expect $\la L(t) \ra$ to
have a have a classical limit. Consider for instance the situation 
where $X$ and $Y$ are larger than $\SL$ and where $T \gg 1/\SL$.
The average boundary lengths will be of order $1/X$ and $1/Y$. But 
for $ 0 \ll t \ll T $ the system has forgotten everything about 
the boundaries and the expectation value of $L(t)$ is, up to 
corrections of order $e^{-2\SL t}$ or $e^{-2\SL (T-t)}$, determined 
by the ground state of the effective Hamiltonian $H_{eff}$
corresponding to $G_\La(X,Y;T)$ (see \cite{al} for details and 
\cite{alnr} for a discussion of various forms of $H_{eff}$. Here we
do not need the explicit expression for $H_{eff}$). 
One finds for this ground state $\la L \ra = 1/\SL$. This picture is confirmed 
by an explicit calculation using eq.\ \rf{2.a7}
as long as $X,Y > \SL$. The system is thus, except for boundary 
effects, entirely determined by the quantum fluctuations of the 
ground state of $H_{eff}$.

We will here be interested in a different and more 
interesting situation where a non-compact 
space-time is obtained as a limit of the compact space-time described 
by \rf{2.a7}. Thus we want to take $T \to \infty$ and at the same 
time also the length of the boundary corresponding to  
proper time $T$ to infinity. 
Since $T \to \infty$ forces $\bX(T,X) \to \SL$ it follows from 
\rf{2.a3} that the only choice of boundary cosmological constant $Y$
independent of $T$ where the length  $\la L(T)\ra_{X,Y,T}$ goes
to infinity for $T \to \infty$ is $Y\equ \mi \SL$ since we have:
\beq\label{2.a8}
\la L(T)\ra_{X,Y,T} = -\frac{1}{G_\La (X,Y;T)} \, 
\frac{\prt G_\La (X,Y;T)}{\prt Y} =  \frac{1}{\bX(T,X)+Y}.
\eeq 

With the choice $Y \equ \mi \SL$ one obtains from \rf{2.a7} in the 
limit $T \to \infty$:
\beq\label{2.a9}
\la L(t) \ra_{X} = \frac{1}{\SL} \; \sinh (2\SL(t+t_0(X))),
\eeq
where $t_0(X)$ is define in eq.\ \rf{2.4}.

We have called $L_2$ the (spatial) length of the boundary corresponding to $T$
and $\la L(t) \ra_X$ the
spatial length of a time-slice at time $t$ in order to
be in accordance with earlier notation \cite{nakayama,al}, 
but starting from a lattice
regularization and taking the continuum limit $L$ is only determined 
up to a constant of proportionality which we fix by 
comparing with a continuum effective
action. In the next section we will show that such a comparison leads
to the identification  of $L$ as $L_{cont}/\pi$ and we are led to the
following 
\beq\label{2.a10}
L_{cont}(t) \equiv \pi \la L(t)\ra_X = \frac{\pi}{\SL} \; 
\sinh (2\SL(t+t_0(X))).
\eeq
Consider the classical surface where the intrinsic geometry is defined
by proper time $t$ and spatial length  $L_{cont}(t)$ of the curve
corresponding to constant $t$. It has the line element
\beq\label{2.a11}
\d s^2 = \d t^2 + \frac{L_{cont}^2}{4\pi^2}\; \d \th^2 =
\d t^2 + \frac{\sinh^2 (2\SL (t+t_0(X)))}{4 \La} \;\d \th^2,
\eeq
where $t \ge 0$ and $t_0(X)$ is a function of the boundary cosmological
constant $X$ at the boundary corresponding to $t \equ 0$ (see eq.\ \rf{2.4}).
What is remarkable about the formula \rf{2.a11} is that the surfaces
for different boundary cosmological constants $X$ can be viewed as
part of the same surface, the Poincare disk with curvature $R= -8\La$,
since $t$ can be continued to $t=\mi t_0 $. The Poincare disk itself is
formally obtained in the limit $X \to \infty$ since an infinite boundary
cosmological constant will contract the boundary to a point.

\section*{The classical effective action}

Consider the non-local ``induced'' action of 2d quantum gravity, first 
introduced by Polyakov \cite{polyakov}
\beq\label{3.1}
S[g]= \int \d t \d x \sqrt{g} \left( \frac{1}{16}
R_g \frac{1}{-\Del_g} R_g +\La \right),
\eeq
where $R$ is the scalar curvature corresponding to the metric
$g$, $t$ denotes ``time'' and $x$ the ``spatial'' coordinate.

Nakayama \cite{nakayama}
analyzed the action \rf{3.1} in proper time gauge assuming
the manifold had the topology of the cylinder with a foliation in
proper time $t$, i.e. the metric was assumed to be of the form:
\beq\label{3.3}
g = \begin{pmatrix} 1&  0 \\ 0 &  \g(t,x)\end{pmatrix}.
\eeq
It was shown that in this gauge the classical dynamics is described
entirely by the following one-dimensional action:
\beq\label{3.4}
S_\k = \int_0^T \d t \left(\frac{\dot{l}^2(t)}{4l(t)}  +
\La l(t)+ \frac{\k}{l}\right),
\eeq
where
\beq\label{3.5}
l(t) = \frac{1}{\pi}\int \d x \sqrt{\g},
\eeq
and where $\k$ is an integration constant coming from solving
for the energy-momentum tensor component $T_{01}=0$ and inserting
the solution  in \rf{3.1}.

Thus $\pi l(t)$ is precisely the length of the spatial curve
corresponding to a constant value of $t$, calculated
in the metric \rf{3.3}. The classical solutions corresponding to action
\rf{3.4} are
\bea
l(t) &=& \frac{\sqrt{\k}}{\SL} \; \sinh 2\SL t,~~~~~~~~~~\k>0~~
\mbox{elliptic case},
\label{3.6a}\\
l(t) &=& \frac{\sqrt{-\k}}{\SL} \;
\cosh 2\SL t,~~~~~~~\k<0~~\mbox{hyperbolic case},
\label{3.6b}\\
l(t) &=& \e^{2\SL t},~~~~~~~~~~~~~~~~~~~~~~~\k=0 ~~\mbox{parabolic case},
\label{3.6c}
\eea
all corresponding to cylinders with constant negative curvature $-8 \La$.

In the elliptic case, where $t$ must be larger than zero,  
there is a conical 
singularity at $t =0$ unless $\k = 1$. For $\k \equ 1$ the geometry is regular
at $t=0$ and this value of $\k$ corresponds precisely to the Poincare 
disk, $t=0$ being the ``center'' of the disk.

Nakayama quantized the actions $S_\k$ for $\k = (m+1)^2$, $m$ a non-negative 
integer, and for $m \equ 0$ he obtained precisely the propagator
obtained by the CDT path integral approach.

\section*{Quantum fluctuations}

In many ways it is more natural to fix the
boundary cosmological constant than to fix the length
of the boundary. However, one pays the price that the
fluctuations of the boundary size are large, in fact of the
order of the average length of the boundary itself \footnote{This
is true also in Liouville quantum theory, the derivation essentially the
same as that given in \rf{5.1}, as is clear from \cite{aagk}.}: 
from \rf{2.a8} we have
\beq\label{5.1}
\la L^2(T) \ra_{X,Y;T} - \la L(T) \ra^2_{X,Y;T} =
-\frac{\prt \la L(T) \ra_{X,Y;T}}{\prt Y} =  \la L(T) \ra^2_{X,Y;T}.
\eeq
Such large fluctuations are also present around
$\la L(t)\ra_{X,Y;T}$ for $t< T$. From this point of view
it is even more remarkable $\la L(t)\ra_{X,Y=- \SL;T=\infty}$ has
such a nice semiclassical interpretation. Let us now by hand fix the
boundary lengths $L_1$ and $L_2$. This is done in the
Hartle-Hawking Euclidean path integral when the geometries $[g]$ are
fixed at the boundaries \cite{hh}. For our one-dimensional boundaries the
geometries at the boundaries are uniquely
fixed by specifying the lengths of the boundaries, and the relation
between the propagator with fixed boundary cosmological constants
and with fixed boundary lengths is given by a Laplace transformation
as shown in eq.\ \rf{2.a5}.
Let us for simplicity analyze the situation where we
take the length $L_1$ of the entrance loop to zero
by taking the boundary cosmological constant $X \to \infty$.
Using the decomposition property \rf{2.a6} one can
calculate the connected ``loop-loop'' correlator for fixed $L_2$ and
$0< t \leq t+\Del < T$
\beq\label{5.a1}
\la L(t)L(t+\Del)\ra^{(c)}_{L_2,T} \equiv
\la L(t+\Del)L(t)\ra_{L_2,T}-\la L(t)\ra \la L(t+\Del)\ra_{L_2,T}.
\eeq
One finds
\bea\label{5.a2}
\la L(t)L(t\pl\Del)\ra^{(c)}_{L_2,T} &=&
\frac{2}{\La}
\frac{\sinh^2 \SL t \sinh^2 \SL (T\mi (t\pl\Del))}{\sinh^2 \SL T}+
 \\
&&
\frac{2L_2}{\SL}
\frac{\sinh^2 \SL t \sinh\SL (t\pl\Del)
\sinh\SL(T \mi(t\pl\Del))}{\sinh^3 \SL T}.\no
\eea
We also note that
\beq\label{5.b2}
\la L(t)\ra_{L_2,T}=
\frac{2}{\SL} \frac{\sinh \SL t \sinh \SL (T\mi t)}{\sinh \SL T}+
L_2\frac{\sinh^2 \SL t}{\sinh^2 \SL T}.
\eeq
For fixed $L_2$ and $T \to \infty$ we obtain
\beq\label{5.a3}
\la L(t)L(t+\Del)\ra^{(c)}_{L_2} =
\frac{1}{2\La} \; \e^{-2\SL \Del } \left(1-\e^{-2\SL t} \right)^2
\eeq
and 
\beq\label{5.b3}
\la L(t)\ra_{L_2}=\frac{1}{\SL}\left( 1-\e^{-2\SL t}\right). 
\eeq
Eqs.\ \rf{5.a3} and \rf{5.b3}  tell us that except for small $t$
we have   $\la L(t)\ra_{L_2}\equ 1/\SL$. The quantum 
fluctuations $\Del L(t)$ of $L(t)$ are defined by 
$(\Del L(t))^2 = \la L(t)L(t)\ra^{(c)}$.
Thus the spatial extension of the universe 
is just quantum size (i.e.\ $1/\SL$, $\La$ being the only coupling
constant) with fluctuations $\Del L(t)$ of the same size.
The time correlation between 
$L(t)$ and $L(t+\Del)$ is also dictated by the scale $1/\SL$, telling 
us that the correlation between spatial elements of size $1/\SL$,
separated in time by $\Del$ falls off exponentially as 
$e^{-2\SL \Del}$ . The above picture is 
precisely what one would expect from the action \rf{2.a}: if we force 
$T$ to be large and choose a  $Y$ such that $\la L_2(T)\ra$ is not 
large, the universe will be a thin tube, ``classically'' of zero 
width, but due to quantum fluctuations of average width $1/\SL$.

A more interesting situation is obtained if we choose $Y = -\SL$,
the special value needed to obtain a non-compact geometry in the limit 
$T\to \infty$. To implement this in a setting where $L_2$ is not 
allowed to fluctuate we fix $L_2(T)$ to the average value \rf{2.a8} for 
$Y\equ \mi \SL$: 
\beq\label{5.2}
L_2(T) = 
\la L(T) \ra_{X,Y= -\SL;T} = \frac{1}{\SL} \; \frac{1}{\coth \SL T -1}.
\eeq
From \rf{5.a2} and \rf{5.b2} we have in the limit $T \to \infty$:
\beq\label{5.3} 
\la L(t) \ra = \frac{1}{\SL} \; \sinh 2\SL t
\eeq
in accordance with \rf{2.a9}, and for the ``loop-loop''-correlator
\beq\label{5.4}
\la L(t+\Del)L(t)\ra^{(c)}=
\frac{2}{\La}\;\sinh^2 \SL t= \frac{1}{\SL} \left( \la L(t)\ra 
-\frac{1}{\SL} \left(1-\e^{-2\SL t}\right)\right).
\eeq

It is seen that the ``loop-loop''-correlator is independent of $\Del$.
In particular we have for $\Del \equ 0$:
\beq\label{5.5}
(\Del L(t))^2 \equiv \la L^2(t)\ra -\la L(t)\ra^2 \sim
\frac{1}{\SL} \la L(t)\ra
\eeq
for  $t \gg 1/\SL$. The interpretation of eq.\ \rf{5.5} is
in accordance with the picture presented below \rf{5.b3}: we can view the
curve of length $L(t)$ as consisting of
$N(t) \approx \SL L(t) \approx e^{2\SL t} $ independently
fluctuating parts of size $1/\SL$ and each with a fluctuation of size
$1/\SL$. Thus the total fluctuation $\Del L(t)$ of $L(t)$ will be of order
$1/\SL \times \sqrt{N(t)}$, i.e.\
\beq\label{5.a5}
\frac{\Del L(t)}{\la L(t)\ra} \sim \frac{1}{\sqrt{\SL \la L(t)\ra}}
\sim \e^{-\SL t},
\eeq
i.e.\ the fluctuation of $L(t)$ around $\la L(t)\ra$ is small for
$t \gg 1/\SL$. In the same way the independence of the ``loop-loop''-correlator
of $\Del$ can be understood as the combined result of $L(t+\Del)$
growing exponentially in length with a factor $e^{2\SL \Del}$ compared to
$L(t)$ and, according to \rf{5.a3},
the correlation of  ``line-elements'' of $L(t)$ and $L(t+\Del)$
decreasing by a factor $e^{-2\SL \Del}$.

\section*{Discussion}

We have described how the CDT quantization of 2d gravity for a special value
of the boundary cosmological constant leads to a non-compact 
(Euclidean) Ads-like space-time of constant negative curvature
dressed with quantum fluctuations. It is possible to achieve this 
non-compact geometry as a limit of a compact geometry as described 
above. In particular the assignment \rf{5.2} leads to a 
simple picture where the fluctuation of $L(t)$ is small 
compared to the average value of $L(t)$. In fact the geometry
can be viewed as that of the Poincare disk with fluctuations 
correlated only over a distance $1/\SL$.

Our construction is similar
to the analysis of $ZZ$-branes appearing as a limit of compact
2d geometries in Liouville quantum gravity \cite{aagk}.
In the CDT case the non-compactness
came when the running   boundary cosmological constant $\bX(T)$ went
to the fixed point $\SL$ for $T \to \infty$.
In the case of Liouville gravity, represented by DT (or equivalently
matrix models), the
non-compactness arose when the running (Liouville) boundary cosmological
constant ${\bX_{liouville}(T)}$ went to
the value where the disk-amplitude $W_\tL(\tX) \equ 0$,
i.e.\ to $\tX \equ \sqrt{\tL}/2$ (see eq.\ \rf{0.0}).
It is the same process in the two cases
since the relation between Liouville gravity
and CDT is well established and summarized by the mapping \cite{ackl}:
\beq\label{6.1}
\frac{X}{\SL} = \sqrt{\frac{2}{3}}\; \sqrt{1+\frac{\tX}{\sqrt{\tL}}},
\eeq
between the coupling constants of the two theories.
The physical interpretation
of this relation is discussed in \cite{ackl,al}: one obtains the CDT model
by chopping away all baby-universes from the Liouville gravity theory,
i.e.\ universes connected to the ``parent-universe'' by a worm-hole
of cut-off scale, and this produces the relation \rf{6.1}
\footnote{The relation \rf{6.1} is similar to the one encountered in
regularized bosonic string theory in dimensions 
$d\geq 2$ \cite{durhuus,adf,ad}:
the world sheet degenerates into so-called branches polymer. The
two-point function of these branched polymers is related to
the ordinary two-point function of the free relativistic particle
by chopping off (i.e.\ integrating out) the branches, just leaving
for each branched polymer connecting two points in target space
one {\it path} connecting the two points. The mass-parameter
of the particle is then related to the corresponding parameter
in the partition function for the branched polymers as
$X/\SL$ to $\tX/\sqrt{\tL}$ in eq.\ \rf{6.1}. }. It is seen
that $X \to \SL$ corresponds precisely to $\tilde{X} \to \sqrt{\tL}/2$.

While the starting point of the CDT quantization was the desire 
to include only Lorentzian, causal geometries in the path integral,
the result \rf{2.a11} shows that after rotation to Euclidean 
signature this prescription is in a natural correspondence with 
the Euclidean Hartle-Hawking no-boundary condition since all of the
geometries \rf{2.a11} have a continuation to $t \equ -t_0$ where 
the space-time is regular.  It would be interesting if this could be 
promoted to a general principle also in higher dimensions. The computer
simulations reported in \cite{ajl5} seems in accordance with 
this possibility.

\section*{Acknowledgment}

J.A.\ and R.J.\ were supported by ``MaPhySto'', 
the Center of Mathematical Physics 
and Stochastics, financed by the 
National Danish Research Foundation. 
All authors acknowledge support by
ENRAGE (European Network on
Random Geometry), a Marie Curie Research Training Network in the
European Community's Sixth Framework Programme, network contract
MRTN-CT-2004-005616.
R.J.\ was supported in part by Polish Ministry of Science and Information
Technologies grant 1P03B04029 (2005-2008)


\begin{thebibliography}{99}


\bibitem{zz}
A.~B.~Zamolodchikov and A.~B.~Zamolodchikov,
{\it Liouville field theory on a pseudosphere,}
arXiv:hep-th/0101152.

\bibitem{martinec}
E.~J.~Martinec,
{\it The annular report on non-critical string theory,}
arXiv:hep-th/0305148.

\bibitem{ss}
N.~Seiberg and D.~Shih,
{\it Branes, rings and matrix models in minimal (super)string theory,}
JHEP { 0402} (2004) 021
[arXiv:hep-th/0312170].
D.~Kutasov, K.~Okuyama, J.~Park, N.~Seiberg and D.~Shih,
{\it Annulus Amplitudes and ZZ Branes in Minimal String Theory,}
JHEP { 0408} (2004) 026
arXiv:hep-th/0406030.

\bibitem{aagk}
J.~Ambj\o rn, S.~Arianos, J.~A.~Gesser and S.~Kawamoto,
{\it The geometry of ZZ-branes,}
Phys.\ Lett.\ B { 599} (2004) 306
[arXiv:hep-th/0406108].

\bibitem{al}
J.~Ambj\o rn and R.~Loll,
{\it Non-perturbative Lorentzian quantum gravity, causality and topology
change,}
Nucl.\ Phys.\ B { 536}, 407 (1998)
[arXiv:hep-th/9805108].

\bibitem{ajl5}
J.~Ambj\o rn, J.~Jurkiewicz and R.~Loll,
{\it Reconstructing the universe},
Phys.\ Rev.\ D\ 72 (2005) 064014 [hep-th/0505154].
{\it Spectral dimension of the universe},
Phys.\ Rev.\ Lett.\ 95 (2005) 171301 [hep-th/0505113].
{\it Semiclassical universe from first principles},
Phys.\ Lett.\ B 607 (2005) 205-213
[hep-th/0411152].
 {\it Emergence of a 4D world from causal quantum gravity,}
  Phys.\ Rev.\ Lett.\  { 93} (2004) 131301
  [arXiv:hep-th/0404156].
{\it Dynamically triangulating Lorentzian quantum gravity},
Nucl.\ Phys.\ B\ 610 (2001) 347-382 [hep-th/0105267].
{\it Non-perturbative 3d Lorentzian quantum gravity},
Phys.\ Rev.\ D\  64 (2001) 044011 [hep-th/0011276].

\bibitem{teitelboim}
Claudio Teitelboim,
{\it Causality versus gauge invariance in quantum gravity and
supergravity},
Phys.\ Rev.\ Lett.\ 50 (1983) 705;
{\it The proper time gauge in quantum theory of gravitation},
Phys.\ Rev.\ D\ 28 (1983) 297.

\bibitem{book}
  J.~Ambj\o rn, B.~Durhuus and T.~Jonsson,
  {\it Quantum geometry. A statistical field theory approach,}
  Cambridge Monogr.\ Math.\ Phys.\  {\bf 1} (1997) 1.

\bibitem{leshouches}
J.~Ambj\o rn:
{\it Quantization of geometry}, in
{\it Fluctuating geometries in statistical mechanics and field theory},
eds. F.\ David, P.\ Ginsparg and J.\ Zinn-Justin,
Elsevier, Amsterdam (1996) 77-193 [hep-th/9411179].

\bibitem{kawai0}
H.~Kawai, N.~Kawamoto, T.~Mogami and Y.~Watabiki,
{\it Transfer matrix formalism for two-dimensional quantum gravity and fractal
structures of space-time,}
Phys.\ Lett.\ B { 306} (1993) 19
[arXiv:hep-th/9302133].

\bibitem{kn}
H.~Aoki, H.~Kawai, J.~Nishimura and A.~Tsuchiya,
{\it Operator Product Expansion in Two-Dimensional Quantum Gravity,}
Nucl.\ Phys.\ B { 474}, 512 (1996)
[arXiv:hep-th/9511117].

\bibitem{gk}
S.~S.~Gubser and I.~R.~Klebanov,
{\it Scaling functions for baby universes in two-dimensional quantum gravity,}
Nucl.\ Phys.\ B {416}, 827 (1994)
[arXiv:hep-th/9310098].

\bibitem{alnr}
  J.~Ambj\o rn, R.~Loll, J.~L.~Nielsen and J.~Rolf,
  {\it Euclidean and Lorentzian quantum gravity:
Lessons from two dimensions,}
  Chaos Solitons Fractals { 10} (1999) 177
  [arXiv:hep-th/9806241].


\bibitem{polyakov}
  A.~M.~Polyakov,
 {\it Quantum Gravity In Two-Dimensions,}
  Mod.\ Phys.\ Lett.\ A { 2} (1987) 893.




\bibitem{nakayama}
  R.~Nakayama,
  {\it 2-D quantum gravity in the proper time gauge,}
  Phys.\ Lett.\ B { 325} (1994) 347
  [arXiv:hep-th/9312158].

\bibitem{hh}
J.B.~Hartle and S.W.~Hawking:
{\it Wave function of the universe,}
Phys.\ Rev.\ D\ 28 (1983) 2960-2975.

\bibitem{durhuus}
  B.~Durhuus, J.~Frohlich and T.~Jonsson,
 {\it Critical Behavior In A Model Of Planar Random Surfaces,}
  Nucl.\ Phys.\ B {\bf 240} (1984) 453
  [Phys.\ Lett.\  {\bf 137B} (1984) 93].

\bibitem{adf}
J.~Ambj\o rn, B.~Durhuus and J.~Frohlich,
{\it Diseases Of Triangulated Random Surface Models, And Possible Cures,}
Nucl.\ Phys.\ B { 257}, 433 (1985).

\bibitem{ad}
J.~Ambj\o rn and B.~Durhuus:
{\it Regularized bosonic strings need extrinsic curvature},
Phys.\ Lett.\ B\ 188 (1987) 253.

\bibitem{ackl}
J.~Ambj\o rn, J.~Correia, C.~Kristjansen and R.~Loll,
{\it On the relation between Euclidean and Lorentzian 2D quantum gravity,}
Phys.\ Lett.\ B { 475} (2000) 24
[arXiv:hep-th/9912267].



\end{thebibliography}
\end{document}